\documentclass[]{siamart171218}

\usepackage{algorithmic}
\usepackage{algorithm}
\usepackage{amsmath}
\usepackage{amssymb}
\usepackage{graphicx}
\usepackage{listings}

\newcommand{\X}{{\bf x}}
\newcommand{\Y}{{\bf y}}
\newcommand{\V}{{\bf v}}

\newcommand{\ass}{{\small\; \mathrel{{:}{=}}\;}}
\newcommand{\inc}{{\small\; \mathrel{{+}{=}}\;}}
\newcommand{\pp}{{\small \mathrel{{+}{+}}}}
\newcommand{\R}{\mathbb{R}}
\newcommand{\Z}{\mathbb{Z}}

\title{Reduction of the Random Access Memory Size in Adjoint Algorithmic Differentiation by Overloading}
\author{Uwe Naumann\thanks{Informatik 12: Software and Tools for Computational Engineering, RWTH Aachen University, Germany. \email{naumann@stce.rwth-aachen.de}}}

\begin{document}

\maketitle


\begin{abstract}
Adjoint algorithmic differentiation by operator and function overloading is 
based on the interpretation of directed acyclic graphs resulting from 
evaluations of numerical simulation programs. The size of the computer 
system memory 
required to store the graph grows proportional to the number of 
floating-point operations executed by the underlying program. It quickly 
exceeds the available memory resources. Naive adjoint algorithmic differentiation often becomes infeasible 
except for relatively simple numerical simulations. 

Access to the data associated with the graph can be classified as sequential and 
random. The latter refers to memory access patterns defined by the adjacency 
relationship between vertices within the graph. Sequentially accessed data can 
be decomposed into blocks. The blocks can be streamed across the system memory hierarchy thus extending the amount of available memory, for example, to 
hard discs. Asynchronous i/o can help to mitigate the increased cost due to
accesses to slower memory. 
Much larger problem instances can thus be solved without resorting 
to technically challenging user intervention such as checkpointing. Randomly 
accessed data should not have to be decomposed. Its block-wise streaming is likely 
to yield a substantial overhead in computational cost due to data 
accesses across blocks. Consequently, the size of the randomly accessed memory
required by an adjoint should be kept minimal in order to eliminate the need 
for decomposition. We propose a combination of dedicated memory for adjoint $L$-values with the exploitation of remainder bandwidth as a possible solution. Test
results indicate significant savings in random access memory size while
preserving overall computational efficiency.
\end{abstract}

\section{Introduction}

Building on prior work in \cite{Naumann2018CSC} we consider a given 
implementation of a differentiable multivariate vector function
$F : \R^n \rightarrow \R^m : \X \mapsto \Y=F(\X)$
over the real numbers $\R$
as a differentiable computer program over floating-point numbers \cite{IEEE754} 
referred to as the {\em primal program}.
Note that differentiability of a function does not imply 
(algorithmic) differentiability of the given implementation. A prime example
is a table lookup of a dedicated value of a differentiable function yielding
a vanishing algorithmic derivative as for example in \lstinline{if (x==0) y=0; else y=sin(x);} 
(Algorithmic) Differentiability of the given implementation
implies differentiability of the function due to the chain rule of 
differentiation.
Algorithmic differentiation (AD) \cite{Griewank2008EDP,Naumann2012TAo} 
yields a variety of implementations of the transformation of the given
differentiable program into a program for evaluating its derivatives. We 
focus on AD by operator and function overloading as supported, for example, by
C++. The following discussion is presented in the context of a serial 
evaluation of $F$ as well as of its (adjoint) derivatives. 
Generalization to parallel
scenarios based on shared \cite{Bischof2008PRM,Forster2014ADo} or distributed 
\cite{Naumann2008AFf,Utke2008TAM} memory architectures
or on massively parallel accelerators \cite{Gremse2015GAA,Moses2021RMA} is 
the subject of ongoing research.

The adjoint \cite{Dunford1988LO} of $F$ becomes
\begin{equation} \label{eqn:bF}
\bar{F} : \R^n \times \R^{1 \times m} \rightarrow \R^{1 \times n} : \quad (\X,\bar{\Y}) \mapsto \bar{\X}=\bar{F}(\X,\bar{\Y}) \equiv \bar{\Y} \cdot F'\; ,
\end{equation}
where 
$$
F'=F'(\X) \equiv \frac{d F}{d \X}(\X) \in \R^{m \times n}
$$
denotes the Jacobian of $F$ at the given point $\X.$ 
Equation~(\ref{eqn:bF}) is evaluated by the {\em adjoint program} resulting 
from the application of adjoint AD to the given implementation of $F$. 

Without loss of generality, the following discussion is based on the 
assumption of a single scope primal. Effects due to allocation and deallocation
of program variables are not accounted for. Additional 
technical issues due to specifics of the programming language as well as due 
to the custom design of the AD tool at hand need to be taken into account when
integrating this paper's ideas into state-of-the-art AD software tools.
Examples throughout this paper as well the reference implementation (see 
Section~\ref{sec:concl})
are written in C++.

Ordered sets 
	$V=\{0,\ldots,n+q-1\}$ (indexes of variables),
	$X=\{0,\ldots,n-1\}$ (indexes of independent input variables)
	and $Y \subseteq V,$ $|Y|=m$ (indexes of dependent output variables)
are induced by the primal program for given values of the inputs.
The primal program is regarded as a composition of {\em elemental functions}
(also: {\em elementals})
$\varphi_j$ evaluated as a {\em single assignment code} (each variable $v_j$ 
is written once as the result of $\varphi_j$)
\begin{equation} \label{eqn:sac}
\begin{split}
        v_i=x_i \qquad \qquad \quad \; \, &\qquad \text{for}~i \in X \; ; \\
	\left .
\begin{split}
v_j&\ass \varphi_{j}\left ((v_i)_{i \prec j} \right ) \\
        y_{k\pp}&=v_j \quad \text{if}~j \in Y
\end{split} \;
	\right \rbrace 
         &\qquad \text{for}~j \in V \setminus Xi~\text{and}~k \ass 0~\text{initially.}
\end{split}
\end{equation}
C++-style incrementation $k\!\pp$ denotes $k\ass k+1.$
Ordered sets are traversed in the given order unless stated otherwise.
For example, $i \in X$  corresponds to ``for $i=0,\ldots,n-1.$''
Following \cite{Griewank2008EDP} we use $i \prec j$ to denote $v_i$ as an argument of $\varphi_j.$
We write $=$ for mathematical equality, $\equiv$ in the sense of ``is defined as'' and $\ass$ to represent assignment as in imperative programming 
languages. 
While all experiments are performed with C++ the conceptual results
are applicable to any imperative programming language offering support for
operator and function overloading.

Equation~(\ref{eqn:sac}) induces a {\em directed acyclic graph} 
(DAG) $G=(V,E)$ with vertices $V=\{0,\ldots,n+q-1\}$ and 
$E \subseteq V \times V.$ Vertices are partitioned into inputs $X=\{0,\ldots,n-1\},$ outputs $Y \in \{0,\ldots,n+q-1\},$ $|Y|=m.$
Local partial derivatives
$d_{j,i} \equiv \frac{\partial v_j}{\partial v_i} \left ((v_l)_{l \prec j} \right )$ 
of the elementals are associated with all edges. 

In the following we propose special treatment of $L$-values
associated with a physical address in the system memory occupied by the primal program.
Such variables can appear on the left-hand side of assignments. The assignment 
operator is invoked on $L$-values only.

The construction of $G=G(\X)$ relies on $\X$ fixing the flow of control in the 
primal program.
Overloading tools for AD use different representations of the DAG which are 
commonly referred to as {\em tapes}.
We separate sequentially and randomly accessed parts of $G$ by storing
them in sequentially accessed memory (SAM) and randomly accessed memory (RAM),
respectively.
$G$ is represented by three arrays ${\bf s} \in \Z^{2 \cdot |V|-|X|+|E|},$ 
${\bf d} \in \R^{|E|},$ and $\bar{\V} \in \R^{|\text{RAM}|}.$
The vector ${\bf s} \subset \text{SAM}$ describes the structure of $G$ starting with the inputs. 
Entries are generated for each elemental function evaluation in form of its
arguments followed by the number of arguments and the result.
The derivatives associated with all edges are stored in ${\bf d} \subset 
\text{SAM}.$ Adjoints are stored in $\bar{\V}=\text{RAM}.$

This internal representation of $G$ is often referred to as a 
{\em gradient tape}. Note that the elementals 
$\varphi_j,$ $j=0,\ldots,n+q-1,$ can represent arbitrary differentiable 
multivariate vector functions. The sole requirement is the availability
of the corresponding adjoint elementals. 
Without loss of generality 
the previously proposed internal representation makes the additional 
assumption that all elementals are scalar functions.

As an alternative to the gradient tape a {\em value tape} would store
the values $v_j$ alongside corresponding operation codes for all 
$\varphi_j,$ $j=0,\ldots,n+q-1.$ Reevaluation of the primal at different 
points requires repeated recording of the gradient tape while value tapes
can be reevaluated by interpretation. An in-depth discussion of the pros and 
cons of gradient vs. value tapes is beyond the scope of this paper. Numerous
technical details and specifics of the given use cases influence the decision 
about the preferred approach.

State of the art implementations of
AD make use of advanced template metaprogramming features
of modern C++ \cite{Hogan2014FRM,Phipps2012EET,Sagebaum2018Etf}. As a result
right-hand sides of assignments or even entire basic blocks can become 
elementals. The entire RAM can be occupied by $L$-values exclusively. The validity of the novel approach to handling of adjoint memory
to be proposed in the following remains unaffected.

Adjoint code generated by source code 
transformation tools such as TAF \cite{Giering1998RfA} or Tapenade 
\cite{Hascoet2013TTA} does not benefit substantially
from the following discussion. It typically uses an exact image (also: 
{\em shadow memory}) of the memory occupied by the primal program for the 
storage of the adjoints. The discussion of potential shortcomings of this 
approach is beyond the scope of this paper.

\paragraph{Example} Let $f: \R \rightarrow \R : 
x \mapsto y=f(x)$ be
implemented in C++ as
\footnote{This example is supposed to illustrate certain aspects of different approaches to the propagation
of adjoints. We do not claim for it to be a reasonable implementation of a practically relevant mathematical model as a computer program; see Section~\ref{sec:cs} for real-world applications.}
\begin{lstlisting}
template<typename T>
void f(std::vector<T>& v) {
  T u;
  for (size_t i=1;i<v.size();i++) {
    u=sin(v[i-1]);
    v[i]=u*u+v[0];
  }
}
\end{lstlisting}
for $\V \in \R^l,$ $v_0=x=1$ and $y=v_{l-1}.$ We set $l=3.$ 
\begin{figure}
\centering
\includegraphics[width=.9\textwidth]{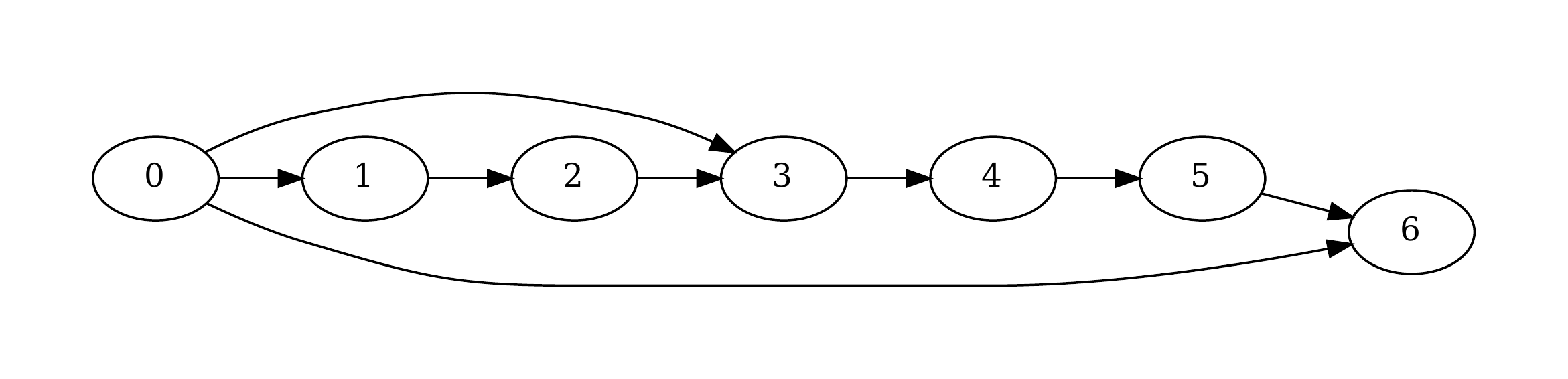}
\caption{DAG} \label{fig:1}
\end{figure}
The corresponding DAG is stored as 
\begin{align*}
{\bf s}^T&=(0~0~1~1~1~1~2~2~0~2~3~3~1~4~4~1~5~5~0~2~6) \in \Z^{21}\\
{\bf d}^T&=(0.54~1.68~1~1~-0.14~1.98~1~1) \in \R^{8} \; .
\end{align*}
The vector ${\bf s}$ is best interpreted in reverse as it is the case in
adjoint AD. Vertex 6 has the 2 
predecessors 0 and 5. Vertex 5 has the single predecessor 4. ... Vertex 1 has the single predecessor 
0. Vertex 0 is the only input. The vector {\bf d} contains the local partial 
derivatives associated with all edges in the order induced by {\bf s}. Edge
$(0,1)$ is labelled with
$$
\frac{\partial v_1}{\partial v_0} =
\frac{\partial \sin(v_0)}{\partial v_0} = \cos(v_0) = \cos(1) \approx 0.54
$$
and so forth.
Floating-point values are rounded to nearest \cite{IEEE754}.
A visualization of the DAG can be found in Figure~\ref{fig:1}. \\
\\
In the following we present three methods for the propagation of adjoints using
different approaches to the allocation of the vector of adjoints 
$\bar{\V} \in \R^{|\text{RAM}|}.$ We aim for $|\text{RAM}| \rightarrow \min.$
Naive adjoint AD records the entire DAG followed by its interpretation as
recalled in Section~\ref{ssec:flat}. Exploitation of bandwidth is likely to
result in a reduction of the adjoint memory requirement as described in 
Section~\ref{ssec:bw}. More substantial improvements can typically be expected 
from the provision of dedicated memory for adjoint $L$-values; see Section~\ref{ssec:lv}.

\section{Flat adjoints by interpretation of the DAG} \label{ssec:flat}

Basic adjoint AD builds $G=(V,E)$ ({\em recording}) followed by allocation of 
$\bar{\V} \in \R^{|V|},$ initialization of 
$\bar{y}_k,$ $k \in Y$ ({\em seeding}), interpretation of $(G,\bar{\V})$ (also: {\em back-propagation}) and extraction of adjoints of the inputs 
$\bar{v}_i,$ $i \in X$ ({\em harvesting}).

For given $\bar{y}_k,$ $k \in Y$ and
setting $\bar{v}_j\ass 0$ for $j \in V \setminus Y,$ back-propagation
amounts to the evaluation of the adjoint program as
\begin{equation} \label{eqn:sac0}
\begin{split}
	\bar{v}_{j_k}&= \bar{y}_k \qquad \quad \; \text{for}~k=0,\ldots,m-1~\text{and}~j_k \in Y\; ; \\
\bar{v}_i&\inc \bar{v}_j \cdot d_{j,i} \quad \text{for}~i \prec j~\text{and}~j=n+q-1,\ldots,n \; ; \\
	\bar{x}_i&=\bar{v}_i \qquad \quad \; \; \forall~i \in X \; .
\end{split}
\end{equation} 
All conditions formulated for Equation~(\ref{eqn:sac}) apply. 
See \cite{Griewank2008EDP} for a proof of correctness of Equation~(\ref{eqn:sac0}).
We use the C++-style notation $\bar{v}_i \inc \bar{v}_j \cdot d_{j,i}$ to abbreviate \mbox{$\bar{v}_i \ass \bar{v}_i + \bar{v}_j \cdot d_{j,i}.$}
RAM of size $|V| \cdot \sigma$ is required, where $\sigma$ denotes the number of bytes occupied by a scalar adjoint, e.g.
$\sigma=8$ for double precision floating-point variables according to the IEEE 754 standard \cite{IEEE754}.

\paragraph{Example} Setting $\bar{y}=1$ yields 
$\bar{\V}^T=(0~0~0~0~0~0~1) \in \R^{7}$
as $|V|=7.$ 
Interpretation makes $\bar{\V}^T$ evolve as
\begin{alignat*}{2}
&(1~0~0~0~0~1~0) &&[\bar{v}_0\inc d_7 \cdot \bar{v}_6;~\bar{v}_5\inc d_6 \cdot \bar{v}_6;~\bar{v}_6\ass 0] \\
&(1~0~0~0~1.98~0~0) &&[\bar{v}_4\inc d_5 \cdot \bar{v}_5;~\bar{v}_5\ass 0] \\
&(1~0~0~-0.27~0~0~0) &&[\bar{v}_3\inc d_4 \cdot \bar{v}_4;~\bar{v}_4\ass 0]\\
&(0.73~0~-0.27~0~0~0~0) &&[\bar{v}_0\inc d_3 \cdot \bar{v}_3;~\bar{v}_2\inc d_2 \cdot \bar{v}_3;~\bar{v}_3\ass 0]\\
&(0.73~-0.46~0~0~0~0~0~) &\quad &[\bar{v}_1\inc d_1 \cdot \bar{v}_2;~\bar{v}_2\ass 0]\\
&(0.48~0~0~0~0~0~0) &&[\bar{v}_0\inc d_0 \cdot \bar{v}_1;~\bar{v}_1\ass 0]
\end{alignat*}
resulting in $\bar{x}=\bar{v}_0=0.48.$ 

Setting $\bar{v}_j \ass 0$ after use in the second line of Equation~(\ref{eqn:sac0}) is actually obsolete for flat adjoints. It becomes essential for ensuring correctness of the alternative methods to be proposed in Sections~\ref{ssec:bw} and \ref{ssec:lv}.

\section{Exploitation of bandwidth} \label{ssec:bw}

The bandwidth $\beta$ of $G=(V,E)$ is defined 
for a given topological 
order of the vertices
as the length $j-i$ of the 
longest edge $(i,j) \in E,$ that is, 
$$\beta \equiv \max_{(i,j) \in E} (j-i) \; .$$ 
The topological order is induced by the sequence of elemental functions 
evaluated by the primal program; see Equation~(\ref{eqn:sac}).
Finding a topological order
which minimizes the bandwidth is known to be NP-complete \cite{Garey1978Crf}.
RAM of size $\max(\beta,n,m)$ is sufficient to evaluate the
primal program and its adjoint, respectively. 

Equation~(\ref{eqn:sac}) can hence be evaluated as
\begin{equation} \label{eqn:sac1}
\begin{split}
	v_i=x_i \qquad \qquad \quad \; \, &\qquad \text{for}~i \in X \; ; \\
\begin{split}
v_{j \% \beta}&\ass \varphi_{j}\left ((v_{i \% \beta})_{i \prec j} \right ) \\
	y_{k\pp}&\ass v_{j\% \beta} \quad \text{if}~j \in Y
\end{split}
	 &\qquad \text{for}~j \in V \setminus X~\text{and}~k\ass 0~\text{initially.} 
\end{split}
\end{equation}
Dedicated outputs are required as the exploitation of bandwidth may yield overwrites; see the explicit assignment in the third line of Equation~(\ref{eqn:sac1}). 
Correctness follows immediately from Equation~(\ref{eqn:sac}) realizing that 
liveness of the $v_j$ can be restricted to the evaluation of all $v_k$ with
$k\leq j+\beta.$

Setting $\bar{v}_j\ass 0$ for $j \in V$ the interpretation of $G$ 
can be performed for given $\bar{\Y}$ in RAM of size $\beta \cdot \sigma$
as
\begin{alignat*}{2}
	\bar{v}_{{(j_k \% \beta)}}&\ass \bar{y}_k &&\text{for}~k=0,\ldots,m-1~\text{and}~j_k \in Y\; ; \\
	\begin{split}
	w &\ass \bar{v}_{(j\%\beta)};~\bar{v}_{(j\%\beta)}\ass 0 \\
	\bar{v}_{(i \% \beta)}&\ass \bar{v}_{(i\%\beta)} + w \cdot d_{j,i}~\text{for}~i \prec j
	\end{split}
		\quad && \text{for}~j=n+q-1,\ldots,n \; ; \\
	\bar{x}_i&=\bar{v}_{(i\%\beta)} &&\text{for}~i=0,\ldots,n-1 \; .
\end{alignat*}
Reuse of RAM locations for distinct variables requires resetting them to zero 
after use in the second equation.
Correctness follows immediately from Equation~(\ref{eqn:sac1}).
\paragraph{Example} Setting $\bar{y}=1$ yields 
$\bar{\V}^T=(1~0~0~0~0~0) \in \R^{6}$
as $\beta=6-0=6.$ 
Interpretation makes $\bar{\V}^T$ evolve as
\begin{alignat*}{2}
	&(1~0~0~0~0~1) &&[w \ass \bar{v}_{0};~\bar{v}_{0} \ass 0;~\bar{v}_0\inc d_7 \cdot w;~\bar{v}_5\inc d_6 \cdot w] \\
&(1~0~0~0~1.98~0) &&[w\ass \bar{v}_{5};~\bar{v}_{5}\ass 0;~\bar{v}_4\inc d_5 \cdot w] \\
&(1 0~0~-0.27~0~0) &&[w\ass \bar{v}_{4};~\bar{v}_{4}\ass 0;~\bar{v}_3\inc d_4 \cdot w]\\
&(0.73~0~-0.27~0~0~0) &\quad &[w\ass \bar{v}_3;~\bar{v}_3\ass 0;~\bar{v}_0\inc d_3 \cdot w;~\bar{v}_2\inc d_2 \cdot w]\\
&(0.73~-0.46~0~0~0~0) &&[w\ass \bar{v}_2;~\bar{v}_2\ass 0;~\bar{v}_1\inc d_1 \cdot w;]\\
&(0.48~0~0~0~0~0) &&[w\ass \bar{v}_{1};~\bar{v}_{1}\ass 0;~\bar{v}_0\inc d_0 \cdot w]
\end{alignat*}
resulting in $\bar{x}=\bar{v}_0=0.48.$ 

The reduction in the size of required RAM is not impressive as the longest edge spans nearly the entire computation. Similar observations can be made for many
practically relevant numerical simulations which is why the exploitation of 
bandwidth alone is typically not enough. {\em Perpetuation} has been proposed
in \cite{Naumann2018CSC} to address this issue. A potentially more powerful
method will be proposed in Section~\ref{ssec:lv}.

The effect of exploiting the bandwidth becomes much more significant for 
simple evolutions of length $l$ defined as 
$$\V=\underset{l~\text{times}}{\underbrace{F(F(\ldots F(\V) \ldots ))}}$$
for $F: \R^n \rightarrow \R^n.$ 
The bandwidth remains bounded by $2 \cdot n$ 
independent of $l.$ The adjoint interpretation of $G$ can be performed in RAM 
of size $2 \cdot n \cdot \sigma.$

%
%
\section{Dedicated adjoint $L$-values} \label{ssec:lv}
Unique locations in system memory are assigned to all $L$-values. A directed 
cyclic graph (DCG) is induced due to potential overwrites. The total size of RAM can 
often be reduced significantly, possibly at the expense of an increase 
in the size of SAM.

Let $V=(L,R),$ $R\equiv V \setminus L$ (remainder) with $X \cup Y \subseteq L.$
Define 
$$\& : V \rightarrow \{-p_L,\ldots,|R|\} : 
i \mapsto 
\begin{cases} 
j \in \{-1,\ldots,-p_L\} & i \in L \\
j \in \{0,\ldots,|R|\} & i \in R 
\end{cases} 
$$
to be injective over $L$ and bijective over $R$ with $p_L$ denoting the
number of distinct $L$-values (with distinct physical addresses) 
in the primal program.

Let $E|_R \equiv \{(i,j) \in E : i,j \in R\}.$
Define the {\em remainder bandwidth} of $G$ as
$$\beta_R=\beta(E|_R) \equiv \max_{(i,j) \in E|_R} \&(j)-\&(i) \; .$$

Let the mapping of DAG vertices $i=0,\ldots,n+q-1$ to RAM with dedicated
$L$-values be defined as
$$
\#(i)\equiv \begin{cases}
\&(i) & \text{if}~i \in L \\
\&(i) \% \beta_R & \text{if}~i \in R \; .
\end{cases}
$$
The primal program can be evaluated in RAM of size $p_L + \beta_R$ as
\begin{alignat*}{2}
v_{\#(i)}&=x_i && \text{for}~i \in X \; ; \\
v_{\#(j)}&\ass \varphi_j\left ((v_{\#(i)})_{i \prec j} \right ) &\quad &\text{for}~j=n,\ldots,n+q-1 \; ;\\
k&\ass 0 \\
y_{k++}&= v_{\#(j_k)} &&\text{for}~j_k \in Y~\text{and}~k \ass 0~\text{initially}\; . 
\end{alignat*}
The corresponding adjoint becomes equal to
\begin{alignat*}{2}
\bar{v}_{\#(j_k)}&= \bar{y}_{k++} &&\text{for}~j_k \in Y~\text{and}~k \ass 0~\text{initially}\; ; \\
	\begin{split}
	w &\ass \bar{v}_{\#(j)};~\bar{v}_{\#(j)} \ass 0 \\
\bar{v}_{\#(i)}&\inc w \cdot d_{j,i} \quad \text{for}~i \prec j
	\end{split}
		\quad && \text{for}~j=n+q-1,\ldots,n \; ; \\
	\bar{x}_i&=\bar{v}_{\#(i)} &&\text{for}~i=0,\ldots,n-1 \; .
\end{alignat*}

\begin{figure}
\centering
\includegraphics[width=.95\textwidth]{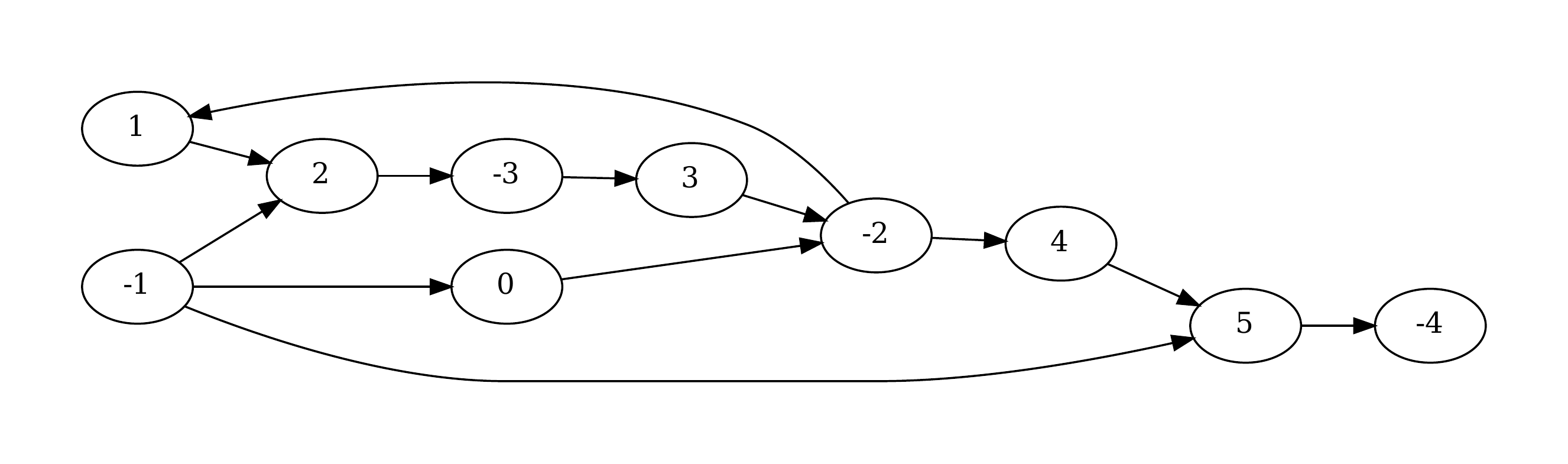}
\caption{DCG} \label{fig:3}
\end{figure}
\paragraph{Example} 
The DCG is stored as 
\begin{align*}
{\bf s}^T&=(-1~-1~1~0~0~1~-2~-2~-2~2~1~1~-1~2~2~2~1~-3~-3~1~3~3~1~-2 \ldots \\
&\qquad \ldots -2~-2~2~4~4~-1~2~5~5~1~-4) \in \Z^{35}\\
{\bf d}^T&=(0.54~1~1.68~1~1~1~-0.14~1~1.98~1~1~1) \in \R^{12}
\end{align*}
A visualization can be found in Figure~\ref{fig:3}. Note that vertex $-2$ corresponds to the program variable \lstinline{u} and is hence visited twice.

Setting $\bar{y}=1$ yields 
$\bar{\V}^T=(0~0~0~1~0) \in \R^{5}$
as $p_L+\beta_R=4+1=5.$ 
The vertices of the DCG are visited as
$$
-4,~5,~-1,~,4,~-2,~3,~-3,~2,~-1,~1,~-2,~0,~-1,
$$
where all non-negative vertices are mapped onto $\bar{v}_4$ while the $L$-values
corresponding to the program variables \lstinline{v[0]}, \lstinline{u}, 
\lstinline{v[1]}, and \lstinline{v[2]} are represented by vertices
$-1,$
$-2,$
$-3,$ and $-4,$ which are mapped onto
$\bar{v}_0,$ 
$\bar{v}_1,$ 
$\bar{v}_2,$ and
$\bar{v}_3,$ respectively. 
Interpretation makes $\bar{\V}^T$ evolve as
\begin{alignat*}{2}
&(0~0~0~0~1) &\quad &[w\ass \bar{v}_3;~\bar{v}_3 \ass 0;~\bar{v}_4\inc d_{11} \cdot w] \\
&(1~0~0~0~1) &&[w\ass \bar{v}_4;~\bar{v}_4\ass 0;~\bar{v}_0 \inc d_{10} \cdot w;~\bar{v}_4\inc d_9 \cdot w] \\
&(1~1.98~0~0~0) &&[w\ass \bar{v}_4;~\bar{v}_4\ass 0;~\bar{v}_1\inc d_8 \cdot w] \\
&(1~0~0~0~1.98) &&[w\ass \bar{v}_1;~\bar{v}_1\ass 0;~\bar{v}_4\inc d_7 \cdot w]\\
&(1~0~-0.27~0~0) &&[w\ass \bar{v}_4;~\bar{v}_4\ass 0;~\bar{v}_2\inc d_6 \cdot w]\\
&(1~0~0~0~-0.27) &&[w\ass \bar{v}_2;~\bar{v}_2\ass 0;~\bar{v}_4\inc d_5 \cdot w]\\
&(0.73~0~0~0~-0.27) &&[w\ass \bar{v}_4;~\bar{v}_4\ass 0;
~\bar{v}_0\inc d_3 \cdot w;
~\bar{v}_4\inc d_4 \cdot w
]\\
&(0.73~-0.46~0~0~0) &&[w\ass \bar{v}_4;~\bar{v}_4\ass 0;~\bar{v}_1\inc d_2 \cdot w]\\
&(0.73~0~0~0~-0.46) &&[w\ass \bar{v}_1;~\bar{v}_1\ass 0;~\bar{v}_4\inc d_1 \cdot w]\\
&(0.48~0~0~0~0) &&[w\ass \bar{v}_4;~\bar{v}_4\ass 0;~\bar{v}_0\inc d_0 \cdot w]
\end{alignat*}
resulting in $\bar{x}=\bar{v}_0=0.48.$ 

\section{Case Studies} \label{sec:cs}
\begin{figure}
\centering
\includegraphics[width=.85\textwidth]{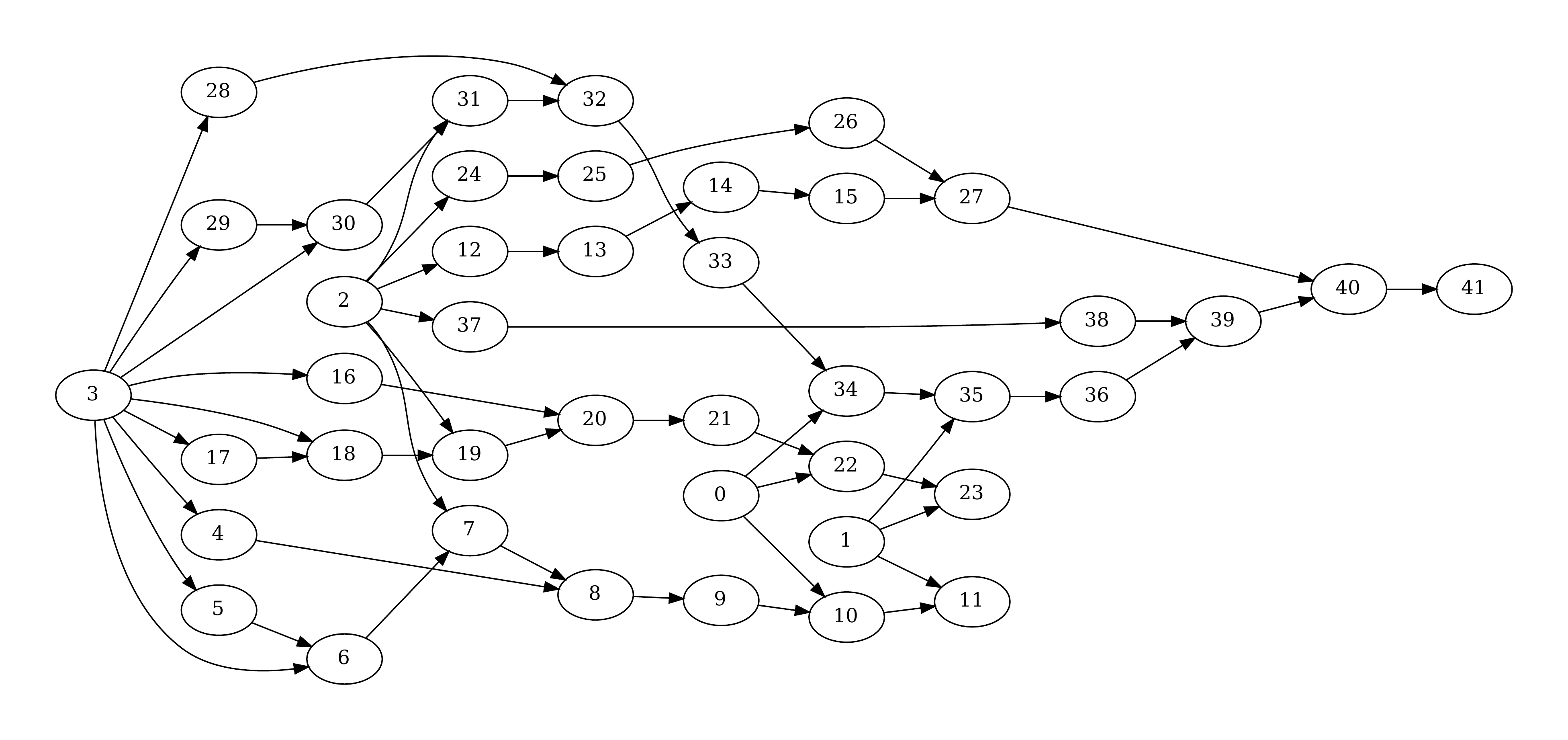}
	\caption{Black-Scholes equation: DAG for three Monte Carlo paths.} \label{fig:5}
\end{figure}

\begin{figure}
\centering
\includegraphics[width=\textwidth]{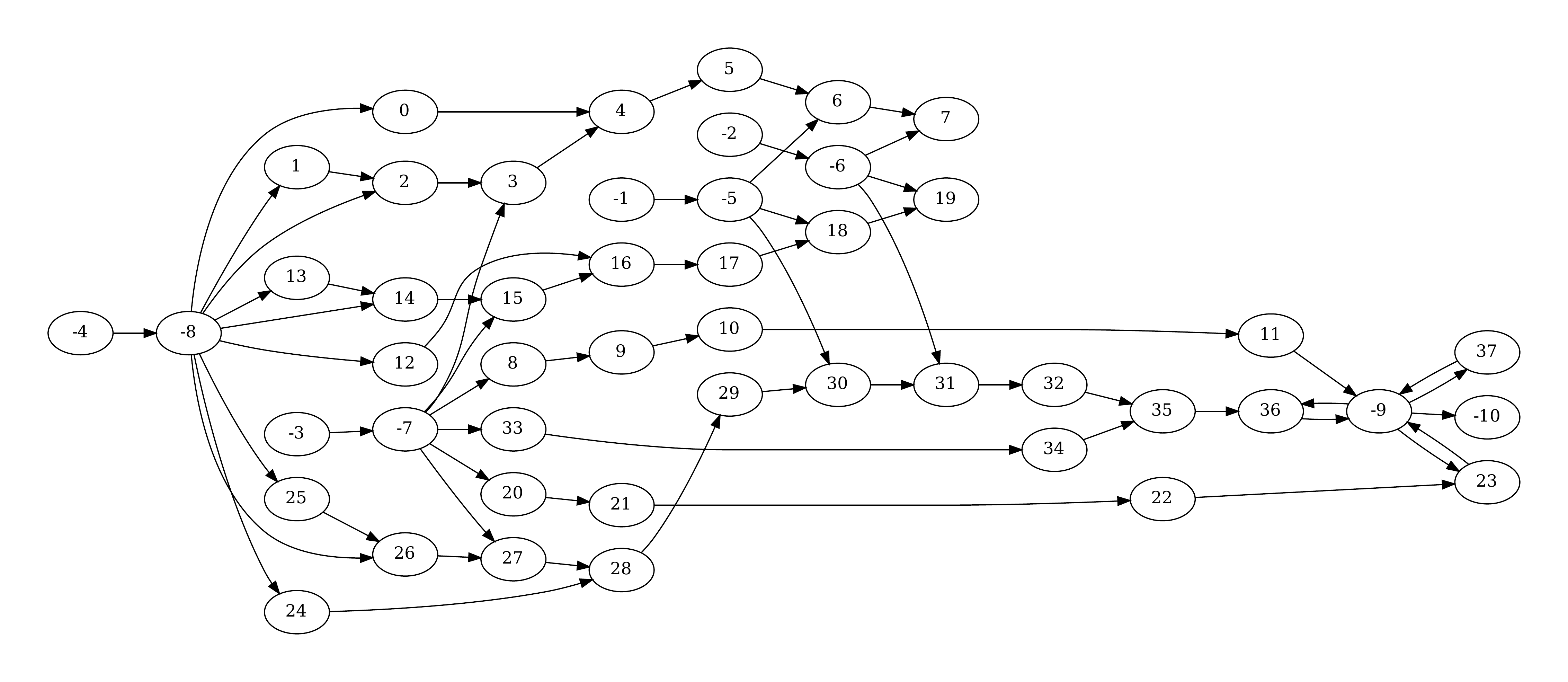}
	\caption{Black-Scholes equation: DCG for three Monte Carlo paths.} \label{fig:6}
\end{figure}

One of the case studies
considers the solution of the Black-Scholes stochastic differential equation
\cite{Black1973TPo} by Monte Carlo simulation \cite{Glas04}. First, we take a closer look at
a simplified scenario running only three paths resulting in the DAG in Figure~\ref{fig:5}. Total flattening yields a RAM requirement of $42 \cdot 8b=336b.$ The 
associated SAM occupies $992b.$
The longest edge is $(2,37)$ yielding a total RAM requirement of $(35+1) \cdot 8b=288b$ under exploitation of bandwidth and with a dedicated adjoint for the single output. The SAM size remains unchanged.
Dedicated adjoint $L$-values lead to the DCG in Figure~\ref{fig:6}.
The longest edge is $(0,4).$ With ten dedicated adjoint $L$-value the total 
RAM requirement becomes equal to $14 \cdot 8b =112b,$ which amounts to 
one third of the RAM required by a total flattening approach due to 
complete mutual independence of the individual paths; see also \cite{Hascoet2002AIC}. 
The SAM requirement is increased slightly to $1172b.$

Test results for a selection of practically relevant problems are listed in 
Tables~\ref{tab:ram} and \ref{tab:sam}. They illustrate the 
potential of dedicating memory to adjoint $L$-values. RAM requirement can be
reduced significantly. We compare RAM sizes resulting from the
approaches discussed in Sections~\ref{ssec:flat}--\ref{ssec:lv} in 
Table~\ref{tab:ram}. 
The following problems are considered:
\begin{itemize}
\item[(a)] Black-Scholes Equation; finite difference scheme on $3 \cdot 10^2 \times 9 \cdot 10^4$ grid \item[(b)] Burgers Equation \cite{Burgers1939Mei}; finite difference scheme with upwind on $10^2 \times 10^4$ grid \item[(c)] LIBOR Market Model \cite{Brace1997TMM}; Monte Carlo simulation with $10^4$ paths 
\item[(d)] Black-Scholes Equation; Monte Carlo simulation with $10^7$ paths.
\end{itemize}
All numerical results were cross-validated and compared with finite difference approximations.
Table~\ref{tab:sam} shows the corresponding SAM sizes. An increase between
seven and up to thirty per cent can be observed. This drawback can be 
mitigated by
asynchronous streaming of the data across the memory hierarchy.
Acceptable slow-down due to the implementation of the additional logic 
or even speed-up due to the reduction in RAM size
was observed as supported by the wall clock time measurements in 
Table~\ref{tab:ram}. All problem sizes allowed for accommodation of both 
RAM and SAM within the system's main memory (approximately 15gb available on
our Intel Xeon workstation running Linux and gcc version 9.4.0).

\begin{table}
\centering
\small
\begin{tabular}{|c|c|c|c|c|}
\hline
	problem & Sec.~\ref{ssec:flat} & Sec.~\ref{ssec:bw} & Sec.~\ref{ssec:lv} \\
\hline
	(a) & 3.383.079.584 (17,4) & 3.383.077.128 (18,1) & 9.736 (17,2) \\
	(b) & 901.360.800 (4,3) & 901.356.056 (4,7) & 133.920.840 (6,0) \\
	(c) & 1.745.200.424 (8,1) & 1.745.199.448 (8,9) & 16.240.664 (9,0) \\
	(d) & 999.989.800 (4,6) & 999.989.744 (5,2) & 112 (5,0) \\
\hline
\end{tabular}
	\caption{RAM requirement in bytes (wall clock times in seconds).}  \label{tab:ram}
\end{table}
\begin{table}
\centering
\small
\begin{tabular}{|c|c|c|c|}
\hline
	problem & Sec.~\ref{ssec:flat} \& Sec.~\ref{ssec:bw} & Sec.~\ref{ssec:lv} \\
\hline
	(a) & 11.306.171.264 & 12.361.174.724 \\
	(b) & 2.863.480.400 & 3.906.400.400 \\
	(c) & 5.232.699.600 & 6.243.799.340 \\
	(d) & 3.279.959.064 & 3.479.959.184 \\
\hline
\end{tabular}
	\caption{SAM requirement in bytes.}  \label{tab:sam}
\end{table}

\section{Conclusion} \label{sec:concl}

The allocation of dedicated memory for adjoint $L$-values combined with
the exploitation of remainder bandwidth yields a substantial reduction
of random access memory requirement in derivative programs obtained
by adjoint AD by overloading. Our reference
implementation can be found on 
\begin{center}
\verb!https://github.com/un110076/dedicated_l_values.git!
\end{center}
allowing for qualitative reproduction
of all results. Its level of sophistication lacks behind state-of-the-art
implementations of adjoint AD by overloading in C++ such as the solutions
offered, for example, by NAG.\footnote{Numerical Algorithms Group Ltd., Oxford, UK; \url{www.nag.com}} It pursues a different purpose by aiming to serve as
an easy-to-follow illustration of the main algorithmic ideas presented in this
paper.

\end{document}